\documentclass[aps,amsmath,asssymb,twocolumn,superscriptaddress,groupedaddress]{revtex4-1}  % for review and submission
\usepackage{graphicx}  % needed for figures
\usepackage{dcolumn}   % needed for some tables
\usepackage{bm}        % for math
\usepackage{amssymb}   % for math
 \usepackage{slashed}
\usepackage{epsf}
\usepackage{hyperref}
\usepackage{enumitem}
\usepackage{lipsum}

\newcommand{\bea}{\begin{eqnarray}}
\newcommand{\eea}{\end{eqnarray}}
\newcommand{\la}{\label}
\newcommand{\be}{\begin{equation}}
\newcommand{\ee}{\end{equation}}
\newcommand{\tr}{\,\mbox{tr}\,}

\begin{document}

\title{Single Spin Asymmetry through QCD Instantons}

\author{Yachao Qian}
\address{Department of Physics and Astronomy,
Stony Brook University,  Stony Brook, NY 11794-3800.}

\author{Ismail Zahed}
\address{Department of Physics and Astronomy,
Stony Brook University,  Stony Brook, NY 11794-3800.}

%\date{\today}

\begin{abstract}
We revisit the effects of QCD instantons in semi-inclusive deep inelastic scattering (SIDIS). We show that large 
single spin asymmetry (SSA) effects can be induced in longitudinally and transversely polarized proton targets.
The results are in agreement with most of the reported data for pion and kaon production. The same effects are found to be important in polarized proton on proton scattering for both charged and chargeless pion productions. The results agree with the reported data in a wide range of $\sqrt{s} = 19.4-200\, {\rm GeV}$. We predict the SSA for $\pi^\pm$ production in $p_\uparrow p$ in the collider range of $\sqrt{s} = 62.4-500\, {\rm GeV}$.  The backward $\pi^{\pm}$ and $\pi^0$  productions for the SSA
in $p_\uparrow p$ collisions are predicted to coincide at large $\sqrt{s}$.
\end{abstract}

\maketitle

%%%%%%%%%%%%%%%%%%%%%%%%%%%%%%
%%%%%%%%%%%%%%%%%%%%%%%%%%%%%%
\section{\label{sec:introduction}introduction}
%%%%%%%%%%%%%%%%%%%%%%%%%%%%%%
%%%%%%%%%%%%%%%%%%%%%%%%%%%%%%
QCD instantons play a central role in the spontaneous breaking of chiral symmetry in the 
QCD vacuum, and may contribute significantly  to the spectroscopic properties of the low 
lying hadrons~\cite{schafer1998,BOOKnowak}.  They may also contribute substantially to semi-hard 
scattering processes by accounting for the soft pomeron 
physics~\cite{shuryak2000,nowak2001,shuryak2004,kharzeev2000,dorokhov2005} 
and possibly gluon saturation at HERA~\cite{ringwald2001, *ringwald1998, *ringwald1994}\cite{ schrempp2003,*schrempp2002a,*schrempp2002b}. 
Recent  lattice gauge simulations may support these claims~\cite{giordano2009,*giordano2011}.

The QCD  instanton  intrinsic spin-color polarization makes them 
ideal for generating non-perturbative and large spin asymmetries in deep inelastic scattering 
using polarized proton  targets~\cite{anselmino1993,kochelev2000, dorokhov2009,ostrovsky2005}. 
Dedicated single spin asymmetry (SSA) experiments in both semi-inclusive deep inelastic scattering
(SIDIS) using $l p_\uparrow \longrightarrow l^\prime \pi X$ by the CLAS and HERMES collaborations
~\cite{hermes2000,hermes2005,hermes2009,clas2010}, as well as polarized proton on proton scattering using $p_\uparrow p \longrightarrow \pi X$
by the STAR and PHENIX collaborations~\cite{star2008,mickey2007,aidala2011,fnal1991}  have unraveled large spin dependent effects.

The large spin asymmetries observed experimentally are triggered by T-odd contributions in the
scattering amplitude.  Since perturbative QCD does not accomodate these effects, it was initially
suggested that these T-odd contributions  are either induced in the initial state (Sivers effect)~\cite{sivers1990,*sivers1991} or in the fragmentation function (Collins effect)~\cite{collins1993,collins1994} thereby preserving the integrety of QCD perturbation theory and factorization. QCD instantons offer a natural mechanism for generating T-odd amplitudes 
that is fully rooted in QCD and beyond perturbation theory~\cite{kochelev2000,dorokhov2009,ostrovsky2005}. 
This approach will be pursued below. 

The organization of the paper is as follows: In section \ref{sec:dip} we revisit and correct the effects of the single instanton 
on the T-odd contribution for the cross section in SIDIS as suggested in~\cite{ostrovsky2005}. We derive the azimuthal
spin asymmetry for longitudinally polarized targets and the Sivers amplitude for transversely polarized targets. By assuming
the instanton parameters to be fixed by their vacuum values, we find good agreement with pion and kaon production
\cite{hermes2000,hermes2005,hermes2009,clas2010}. In section \ref{sec:ppcollision} we extend our analysis to $p_\uparrow p$ collisions and compare to the pion production results reported by STAR and PHENIX~\cite{star2008,mickey2007,aidala2011,fnal1991}. Our results agree with the experimentally measured SSA for $\pi^0$
production in a wide range of $\sqrt{s}$. We predict the SSA for both forward and backward
charged pion in $p_\uparrow p$ collisions for a broad range of collider energies. Remarkably, $\pi^0$ and $\pi^\pm$
productions are found to coincide backward at large $\sqrt{s}$. Our conclusions are in section~\ref{sec:summary}.
Some pertinent calculational details can be found in the Appendices.

%%%%%%%%%%%%%%%%%%%%%%%%%%%%%%
\section{\label{sec:dip}Semi-Inclusive Deep Inealstic Scattering}
%%%%%%%%%%%%%%%%%%%%%%%%%%%%%%
To set up the notations for the semi-inclusive processes in deep inelastic scattering, we consider a proton at
rest in the LAB frame with either longitudinal or transverse polarization as depicted in Fig.~\ref{CARTOONDIS}. 
The incoming and outgoing leptons are unpolarized. The polarization of the target proton in relation to the DIS
kinematics is shown in Fig.~\ref{dip3dgraph} .  Throughout, the spin dependent asymmetries will be evaluated 
at the partonic level. Their conversion to the hadronic level will follow the qualitative arguments presented in~\cite{kochelev2000,dorokhov2009,ostrovsky2005}. 

\subsection{One Instanton Contribution}

Generically, the spin averaged leptonic tensor reads

\begin{figure}
\includegraphics[height=50mm]{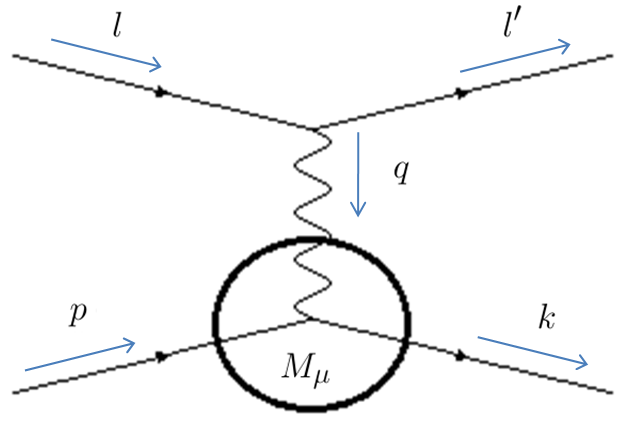}
\caption{ $l$ and $l^\prime$ denote the momentum for the incoming and outgoing lepton. $p$ and $k$ are the momenta of the incoming and outgoing quark. The lepton and the quark exchange one photon in the single instanton background. }
\label{CARTOONDIS}
\end{figure}

\bea\la{leptontensor}
L^{\mu \nu} = \frac{1}{2} \tr[\slashed{l}^\prime \gamma_\mu \slashed{l} \gamma_\nu]
\label{LEPTONIC}
\eea
while the color averaged hadronic tensor in the one instanton background reads

\bea\label{hadrontensor}
W_{\mu \nu} = \sum_{color} \frac{1}{2} \tr[\slashed{k} M_\mu \slashed{p}(1 + \gamma_5 \slashed{s}) M_\nu] 
\label{HADRONIC}
\eea
with the constituent vertex 

\be
M_\mu = \gamma_\mu + M_\mu^{(1)}
\ee
that includes both the perturbative $\gamma_\mu$  and the non-perturbative insertion $M_\mu^{(1)}$. 

In general, $W_{\mu \nu}^{(1)}$ admits all possible tensor structures compatible with gauge and parity invariance,
for a fixed proton spin $s^\mu$. For the Sivers effect to be considered below, there are only {\it two} tensor 
combinations of interest
\be
p_\mu \epsilon_{\nu a b c}s^a k^b  p^c \qquad{\rm and}\qquad q_\mu \epsilon_{\nu a b c}s^a k^b  p^c \nonumber
\ee
The last tensor combination contracts to zero leptonically.
%$ L^{\mu \nu}  q_\mu \epsilon_{\nu a b c}s^a k^b  p^c = 0$. 
Therefore  $W_{\mu \nu}^{(1)}$ can only support the first tensor combination $ p_\mu \epsilon_{\nu a b c}s^a k^b  p^c $. 
This tensor structure is due to the chirality flip vertex $M_\mu^{(1)}$ induced by a single 
instanton. It is at the origin of the SSA in the hard scattering amplitudes to be derived below.
In Appendix \ref{vertex} we detail its derivation following the
original arguments in~\cite{moch1997, ostrovsky2005}

\bea\la{dipm1}
M_\mu^{(1)} =  i \frac{4 \pi^2 \rho^2}{\lambda Q^2}[ \gamma_\mu \slashed{k} + \slashed{p} \gamma_\mu] (1 - f(\rho Q)) 
\label{1VERTEX}
\eea
with $f(a) = a K_1(a)$ and after color averaging. Here $\rho$ is the instanton size and $\lambda$ the renormalized virtuality of
the instanton quark zero mode. Both of these parameters will be discussed below using the QCD instanton vacuum.

The leading non-perturbative instanton contribution to (\ref{CROSS})
is a cross contribution in the hadronic tensor (\ref{HADRONIC}) after inserting the one-instanton vertex (\ref{1VERTEX})

\bea
\label{WXX}
W_{\mu \nu}^{(1)} &=& \frac{1}{2} \tr[\slashed{k} M_\mu^{(1)} \slashed{p} \gamma_5 \slashed{s} \gamma_\nu] + (\mu \leftrightarrow \nu) \\
 &=&   \frac{16 \pi^2 \rho^2}{\lambda Q^2} (1 - f(\rho Q)) (p+k)_{\{ \mu} \epsilon_{\nu \} a b c}s^a k^b  p^c   \nonumber
\eea
where the short notation $(\cdots)_{\{ \mu} \epsilon_{\nu \} a b c} \equiv (\cdots)_\mu \epsilon_{\nu a b c}  + (\cdots)_\nu \epsilon_{\mu a b c}$
is used.    If we set $p=xP$ and $k={K}/{z}$ and note that $p+k = 2 p + q$, then (\ref{WXX}) simplifies
\be\la{hadrontensor2}
 W_{\mu \nu}^{(1)} =   \frac{2^5 \pi^2 \rho^2}{\lambda Q^2} \frac{x^2}{z} (1 - f(\rho Q)) (P + \frac{q}{2x})_{ \{ \mu} \epsilon_{ \nu \} a b c}s^a K^b  P^c 
\ee
Combining (\ref{leptontensor}) and (\ref{hadrontensor2}) yields
\bea\la{lwdip}
W_{\mu \nu}^{(1)}L^{\mu \nu} =&& \frac{2^7 \pi^2 \rho^2}{ \lambda Q^2} \frac{x^2}{z} (1 - f(\rho Q))M \\
&& \times [  E  \epsilon_{\nu a b c} l^{\prime \nu} s^a K^b  P^c +   E^\prime   \epsilon_{\nu a b c} l^{ \nu} s^a K^b  P^c ]  \nonumber\\ \nonumber
\eea
where $E$ ($E'$) is the energy of the incoming (outgoing)  (anti)electron.

The normalized lepton-hadron cross section of Fig.\ref{CARTOONDIS} follows in the form
\bea\la{crosssection}
\frac{d \sigma}{dx dy dz d\phi} = y \frac{\alpha^2}{  Q^6} L^{\mu \nu} W_{\mu \nu} \sum_i e_i^2 f_i (x, Q^2) D_i (z)
\label{CROSS}
\eea
with $y = P \cdot q / P \cdot l $,  where $e_i$ is the i-parton electric charge, $f_i$ its momentum fraction distribution and $D_i$ its framentation function.  The perturbative contribution follows by setting $M_\mu\approx \gamma_\mu$, leading to

\be\la{zeroorder}
\frac{d^{(0)} \sigma}{dx dy dz d\phi} =2 N_c  \frac{\alpha^2}{Q^2} \frac{1+ (1-y)^2}{y} \sum_i e_i^2 f_i (x, Q^2) D_i (z)
\ee
where $N_c=3$ is the number of colors. The sum is over the electrically charged quarks.

\begin{figure}[b]
\includegraphics[height=43mm]{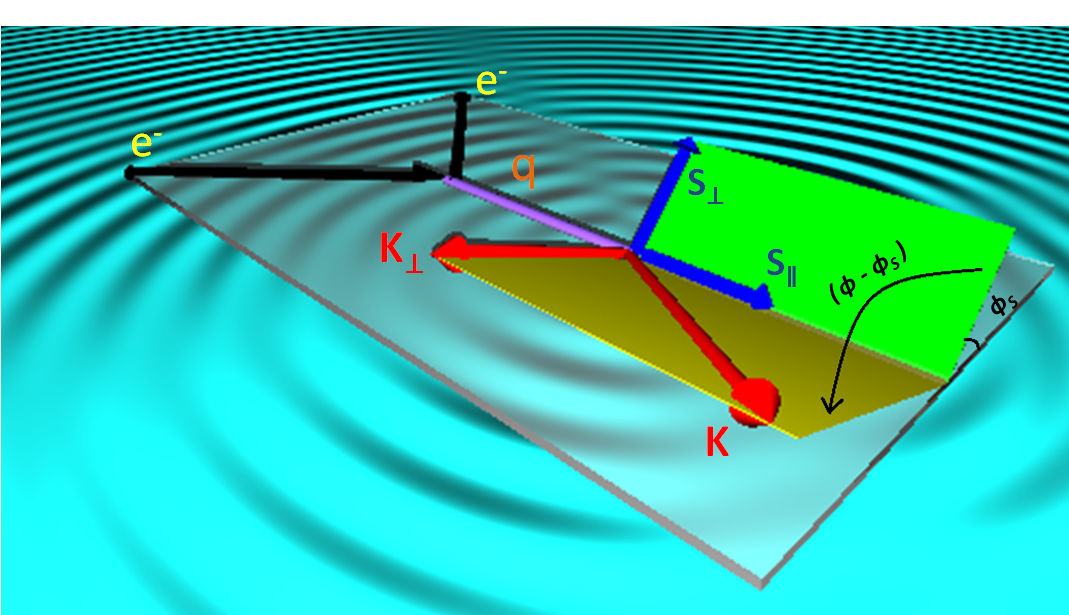}
\caption{\la{dip3dgraph}   The lepton and photon are in the same plane. The angle between the transversely polarized spin $s_\perp$ and this plane is $\phi_s$. The angle between the transversely spatial  momentum $K_\perp$ of the outgoing pion and the plane is $\phi$. The convention for azimuthal angle is consistent with experiments.}
\end{figure}

The leading instanton contribution to the total cross section (\ref{CROSS}) follows by inserting (\ref{lwdip}) into (\ref{HADRONIC}).
For the target proton polarization set up of Fig.~\ref{dip3dgraph}, the longitudinal ($||$) and transverse ($\perp$) cross sections
contribute to the total cross section as

\begin{widetext}
\bea\label{SIGMA1}
\frac{d^{(1)} \sigma}{dx dy dz d\phi} &=& \frac{\alpha^2}{y Q^2}  \sum_q \frac{64 \pi^2 \rho^2   e_q^2  D_q(z)}{ \lambda Q } \frac{K_\bot}{z Q} \frac{M}{Q}  (1 - f(\rho Q)) \nonumber\\
&& \times   [ \Delta_\parallel q_qx (2-y) \sqrt{\frac{1-y-\frac{M^2}{Q^2} x^2 y^2 }{1 + 4 \frac{M^2}{Q^2}x^2}} \sin\phi + \Delta_\perp q_q \frac{Q}{|M|} \frac{1-y-x^2 y^2 \frac{M^2}{Q^2}}{\sqrt{1 + 4 \frac{M^2}{Q^2}x^2}}  \sin(\phi - \phi_s)   ] 
\eea
\end{widetext}
where $\Delta_\parallel q_q (x, Q^2) = s_\parallel f_{q}(x, Q^2)$ is the spin polarized distribution function for the quark in the longitudinally polarized proton, and $\Delta_\perp q_q (x, Q^2) = s_\perp f_{q}(x) $ is the spin polarized distribution function for the quark in the transversely polarized proton.  The overall
sign in (\ref{SIGMA1}) is tied with the conventional sign of the proton mass $M$.
For further comparison with experiment, we will assume (see HERMES~\cite{hermes1998})
that $ \Delta_\parallel q_q (x, Q^2) = \Delta_\perp q_q (x, Q^2)$ and define the spin structure function

\be
  g_1(x,Q^2) = \frac{1}{2} \sum_q e_q^2 (\Delta q_q (x,Q^2) + \Delta \bar{q}_q (x,Q^2))  
\ee

Since we are only interested in the SSA in hard scattering processes, we set $D_q (z) = 1$.  A comparison of (\ref{zeroorder}) with (\ref{SIGMA1}), yields for the longitudinal spin asymmetry (part proportional to ${\rm sin}\phi$)

\bea\la{dipsin}
A^{{\rm sin}\phi}_{UL} &=&  \frac{32 \pi^2 \rho^2}{N_c Q \lambda}  (1 - f(\rho Q))\frac{ K_\bot}{zQ}\frac{M}{Q} \nonumber\\
&& \times 
  \frac{ x (2-y)  }{ 1 + (1-y)^2}\sqrt{\frac{1-y-\frac{M^2}{Q^2} x^2 y^2 }{1 + 4 \frac{M^2}{Q^2}x^2}}  \frac{g_1}{F_1} 
\eea
with $F_1(x) = \frac{1}{2} \sum_{q} e_q^2 f_q(x)$ . The same comparison, yields for the transverse spin asymmetry
(part proportional to ${\rm sin}(\phi-\phi_s)$)

\bea\la{dipsindiff}
A^{{\rm sin}(\phi - \phi_s)}_{UT} &=& \frac{32 \pi^2 \rho^2}{N_c Q \lambda}  (1 - f(\rho Q)) \frac{ K_\bot}{zQ} \frac{M}{|M|}
 \nonumber\\
&& \times \frac{1 }{ 1 + (1-y)^2} \frac{1-y-x^2 y^2 \frac{M^2}{Q^2}}{\sqrt{1 + 4 \frac{M^2}{Q^2}x^2}}      \frac{g_1}{F_1} 
\eea

This is usually referred to as the Sivers contribution.

The longitudinal to transverse ratio simplifies

\be\la{rationew}
\frac{A^{{\rm sin}\phi}_{UL}}{A^{\sin (\phi - \phi_s)}_{UT}} = \frac{|M|}{Q} x \frac{2 - y}{\sqrt{1-y-x^2 y^2 \frac{M^2}{Q^2}}}
\ee
as the details of the instanton form factor and the partonic distributions drop out.  In the hard scattering limit
with large $Q^2$, i.e. $\frac{M^2}{Q^2} \ll  1$,  (\ref{rationew}) reads

\be
\frac{A^{sin\phi}_{UL}}{A^{\sin (\phi - \phi_s)}_{UT}} =  \frac{|M|}{Q} x \frac{2 - y}{\sqrt{1-y}}
\ee

\subsection{Results versus Data}

To compare with SIDIS experiments we need to set up the instanton parameters.  The instanton size will be set to its 
mean value in the QCD vacuum, $\rho\approx 1/3$ fm or $1.7\,/{\rm GeV}$~\cite{schafer1998,BOOKnowak}. The mean instanton 
quark zero mode virtuality is tied to the light quark condensate $\chi_{uu}$ through $\bar{\lambda}=1/\chi_{uu}$. For
two flavors $\chi_{uu}\approx (200\,{\rm MeV})^3$, so that $\bar{\lambda}\approx 1/(0.2\,{\rm GeV})^3$. Thus
$16\pi^2\rho^2/N_c\lambda\approx 1.22\,{\rm GeV}$ in (\ref{dipsin}-\ref{dipsindiff}).

We recall that in the recent CLAS experiment~\cite{clas2010} the kinematical ranges are: 
$Q^2\approx (0.9 - 5.4)\,{\rm GeV}^2$, $x\approx (0.12 - 0.48)$, $z\approx (0.4 - 0.7)$ and $K_\perp \approx (0-1.12)\,{\rm GeV}$. 
For comparison, we use the mean values $<z>\approx 0.55$ and $<K_\perp> \approx 0.56\,{\rm GeV}$. For the spin structure function
we use $g_1^p/F_1^p\approx 0.354$ for $(x,Q^2)\approx (0.274, 3.35\,{\rm GeV}^2)$ from the tabulated values in~\cite{hermes1998}.
This choice of $(x,Q^2)$ agrees with the averaged values of $x$ and $Q^2$ as covered by the CLAS experiment~\cite{clas2010}.
The final result is averaged over the full measurement range of $y\approx (0 - 0.85)$. Fig.~\ref{aul} shows our result 
for the longitunally polarized spin asymmetry versus azimuth. The data is from the CLAS experiment~\cite{clas2010}.

\begin{figure}
\includegraphics[height=52mm]{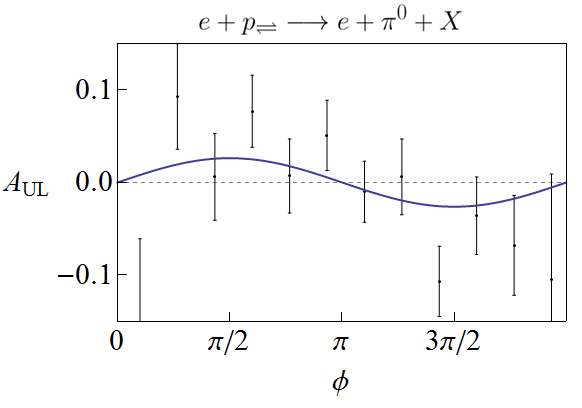}
\caption{Longitudinally polarized spin asymmetry versus azimuth $\phi$. The data are from~\cite{clas2010}.}
\la{aul}
\end{figure}

The kinematical ranges in the longitudinally polarized HERMES experiment~\cite{hermes2000} are:
$<K_\perp> = 0.44\,{\rm GeV}$, $<z> = 0.48$ and $0.2<y<0.85$. As $<Q^2>$ is set
for each data point, we parametrize by $\sqrt{<Q^2>} = 4.3747 <x>^{0.4099}$  ($R^2 = 0.998$). 
This kinematical set up is similar to the one adopted in~\cite{ostrovsky2005} and will facilitate
the comparison. The spin structure function in~\cite{hermes2000} is $g_1^p/F_1^p = 0.9811<x>^{0.7366}$ ($R^2 = 0.98$).
In Fig.~\ref{aul2} we show our results (solid line) 
for the longitudinal spin asymmetry (\ref{dipsin}) versus the data. The dashed line is the result in~\cite{ostrovsky2005}. The
differences stem from corrections to the analysis in~\cite{ostrovsky2005}.  

\begin{figure}
\includegraphics[height=43mm]{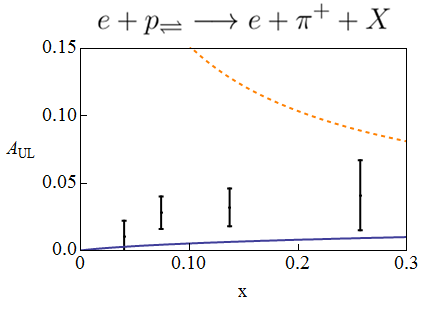}
\caption{\la{aul2} Longitudinally polarized spin asymmetry~\cite{hermes2000}. Our result (\ref{dipsin}) (solid line) and the result in~\cite{ostrovsky2005} (dashed line). }
\end{figure}

The kinematical ranges for the transversely polarized HERMES experiment~\cite{hermes2005} are:
$<x>=0.09$,  $<y>=0.54$, $<z>=0.36$ , $<K_\bot> = 0.41\,{\rm GeV}$ and $<Q^2> = 2.41\,{\rm GeV}^2$.
In Fig.~\ref{aut} we show our results (solid line) for the transversely polarized spin asymmetry (\ref{dipsindiff}) versus
data. Our results correct minimally those reported in~\cite{ostrovsky2005}. The one instanton contribution to the spin 
asymmetries for the transversely polarized proton are larger than the reported data~\cite{hermes2005}.

\begin{figure}
\includegraphics[height=38mm]{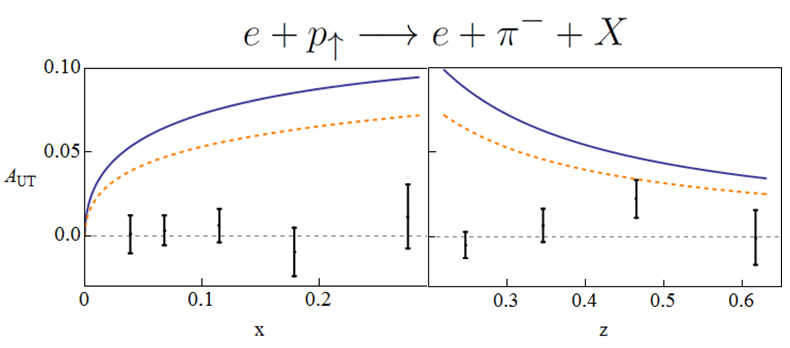}
\caption{\la{aut} Transversely polarized spin asymmetry~\cite{hermes2005} for $\pi^-$ production. $Q^2$ is prametrized as $\sqrt{<Q^2>} = 2.7408 <x>^{0.2793}$ ($R^2 = 0.9878$)~\cite{hermes1998} for the left graph, and $<Q^2> =2.41 GeV^2$~\cite{hermes2005} for the right graph. Our result (solid line) and the result in~\cite{ostrovsky2005} (dashed line).}
\end{figure}

A better comparison with the data can be made using the recently reported HERMES results~\cite{hermes2009}. 
While the use of the average kinematics above is well motivated, a direct probing of the dependence of the transverse
spin asymmetry on $x,z,K_\perp$ is even better.  For instance, take the $x$ dependent asymmetry of $\pi^-$ with
the empirical parametrizations to fit the reported kinematics from HERMES~\cite{hermes2009}:
 $<y> = -95.737 x^3 + 52.459 x^2 - 9.0816 x + 0.9495$, $<z>= 15.67 x^3 - 8.8459 x^2 + 1.5193 x + 0.2884$, $<K_\bot> = 665.15 x^4 - 444.02 x^3 + 105.99 x^2 - 10.843 x + 0.7502 \,({\rm GeV})$ , and $<Q^2> = 20.371 x + 0.4998 ({\rm GeV^2})$. $R^2$ is above 0.97 for all the parametrizations.  In Fig.~\ref{autnew} we compare our results (solid line) for the transverse polarization to the recent data
~\cite{hermes2009}.  The instanton induced asymmetry is in overall agreement for most $\pi^+$, $\pi^0$, $\pi^-$, $K^+$ and $K^-$ 
production.

\begin{figure*}
\includegraphics[width = 120mm]{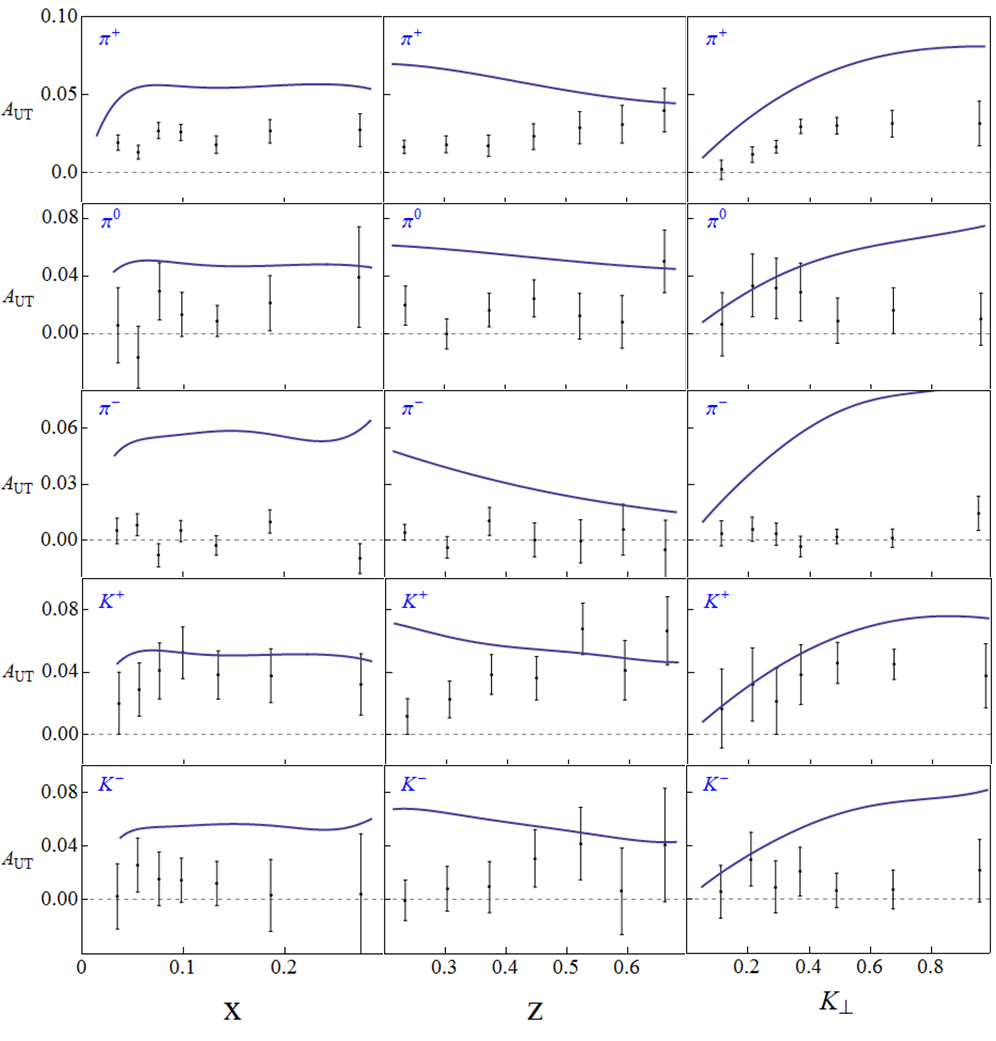}
\caption{\la{autnew} Transversly polarized spin asymmetry (solid line) versus data~\cite{hermes2009}.}
\end{figure*}

%%%%%%%%%%%%%%%%%%%%%%%%%%%%%%%%
\section{\label{sec:ppcollision}$P_\uparrow P$ Collision}
%%%%%%%%%%%%%%%%%%%%%%%%%%%%%%%

The instanton mechanism used to generate T-odd amplitudes in SIDIS as detailed in~\ref{sec:dip} can be extended to 
$p_\uparrow p$ colision. In the CM frame the target proton is transversely polarized, while the projectile proton is unpolarized.
As we are interested in the Sivers effect arising from a hard scattering process, we will only consider one instanton insertion
tagged to a single gluon exchange. The $\pi^+$   productions are displayed in Fig.~\ref{piplus1} and  Fig.~\ref{piplus2}.  The valence quark from the polarized proton exchanges one gluon with the valence quark from the unpolarized proton. The outgoing up-quark with momentum $k$ turns to the forward $\pi^+$.

Throughout, the proton will be viewed as 3 constitutive quarks with the gluons omitted. This simplified partonic description
is motivated by the enhanced role played by the quark zero modes in the instanton background, and will help parallell the 
polarized SIDIS analysis with the inclusive hadron production in polarized $p_\uparrow p$. The translation from partonic to
hadronic results will follow the SIDIS analysis.  The inclusion of constitutive gluons will be addressed elsewhere.

\begin{figure}
\includegraphics[height=75mm]{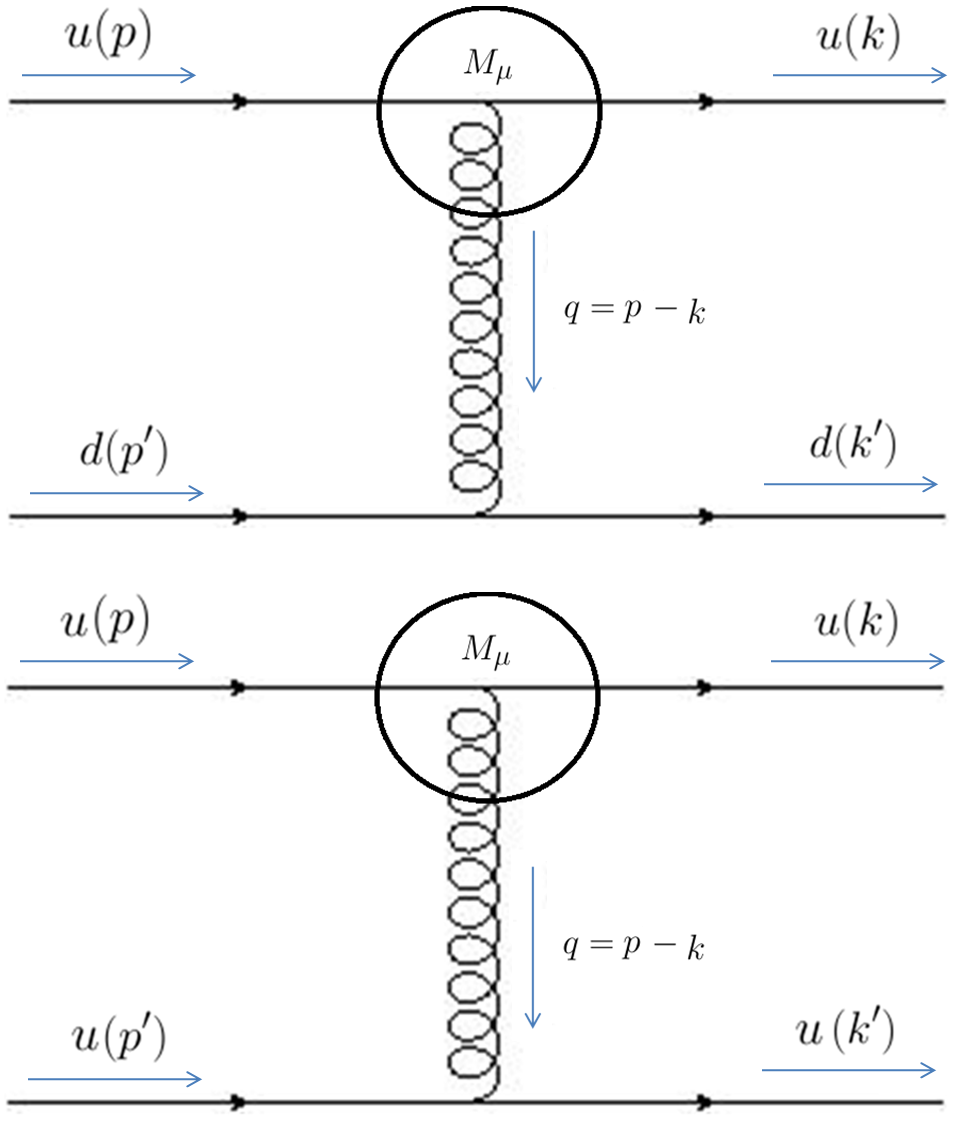}
\caption{\la{piplus1} The up-quark from the transversly polarized proton and the quark from the unpolarized proton exchanges one gluon in the single instanton background (t - channel) . }
\end{figure}

\begin{figure}
\includegraphics[height=78mm]{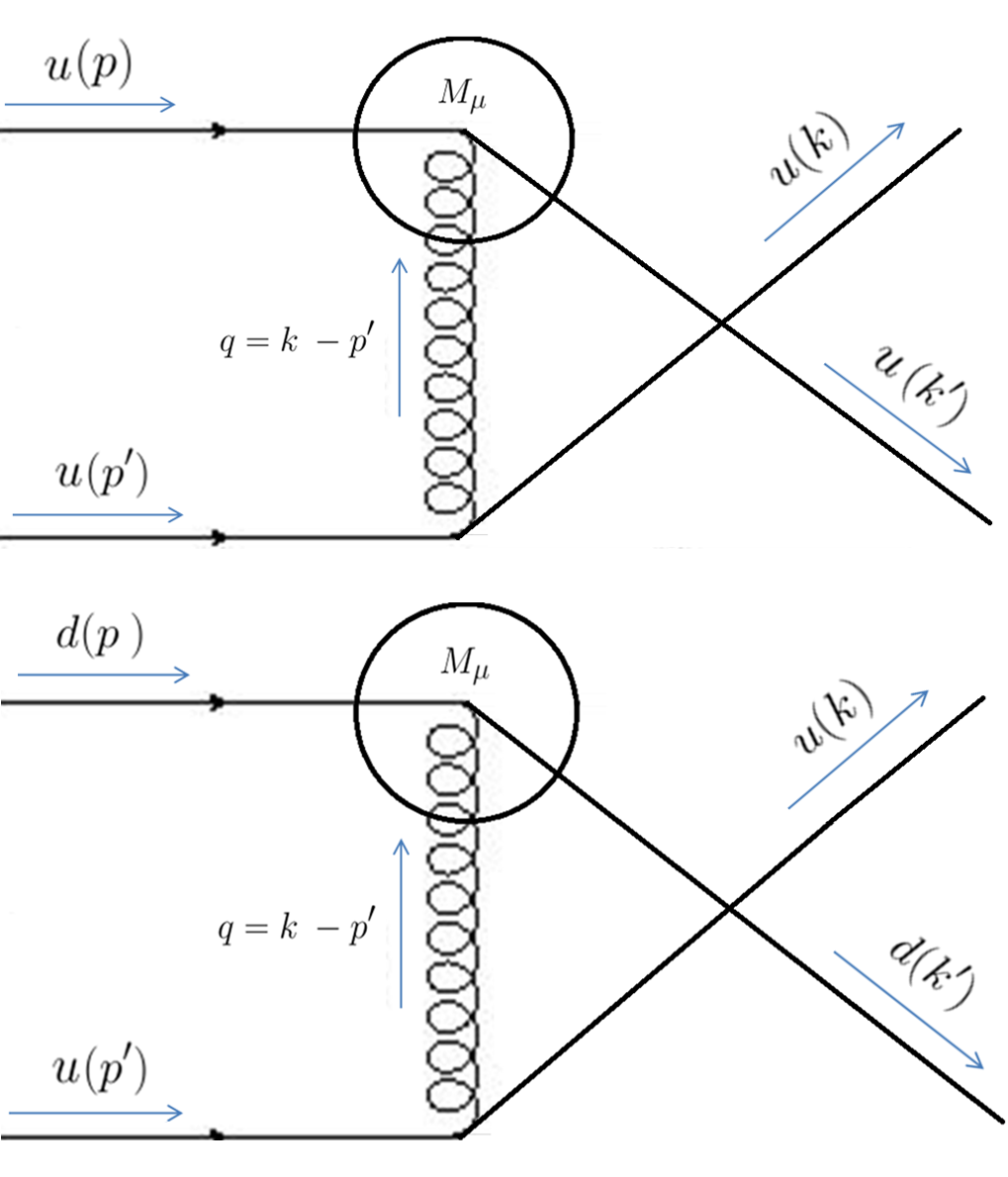}
\caption{\la{piplus2} The quark from the transversly polarized proton and the up-quark from the unpolarized proton exchanges one gluon in the single instanton background (u - channel). }
\end{figure}

The cross section arising from the process in Fig.~\ref{piplus1} reads
\be\la{ppcollision}
d \sigma \sim  \sum_{color} \frac{1}{q^4} \tr [\slashed{k} M_{\mu, a} \slashed{p} (1+ \gamma_5 \slashed{s}) M_{\nu,b} ] \tr [\slashed{k}^\prime M^{\mu,a} \slashed{p}^\prime M^{\nu,b} ]
\ee
As we are only interested in the spin dependent asymmetry, the overall coefficients in (\ref{ppcollision}) will not be needed.
Similarly to SIDIS, we set

\bea
M_{\mu ,a} =  i g_s (\gamma_\mu t_a + M_{\mu , a}^{(1)} )
\eea
with the Gell-Mann color matrix $t_a$ satisfying $\tr[t_a t_b] = \frac{1}{2}\delta_{a b}$.  After a short algebra we have

\bea\la{ppratio1}
d^{(1)}\sigma &\sim& \frac{g_s^4}{q^4} \sum_{color} \tr [( M^{(1)}_{\mu,a} t_b \delta^{a b})  \slashed{p} \gamma_5 \slashed{s} \gamma_\nu \slashed{k}] \tr [\gamma^\mu \slashed{p}^\prime \gamma^\nu \slashed{k}^\prime] \nonumber\\
d^{(0)}\sigma  &\sim&  2 \frac{g_s^4}{q^4} \tr [\gamma_\mu \slashed{p}  \gamma_\nu \slashed{k}] \tr [\gamma^\mu \slashed{p}^\prime \gamma^\nu \slashed{k}^\prime]
\eea

 In Appendix~\ref{ppvertex} we detail the calculation for the instanton induced vertex. In short

\be\la{ppvertexresult}
\sum_{color} M^{(1)}_{\mu,a} t_b \delta^{a b} = -i  (\gamma_\mu \slashed{k} + \slashed{p} \gamma_\mu)   \Psi(\rho |q|) 
\ee
with 
\be
\Psi(a)  \equiv \frac{ 2 \pi^2 \rho^4}{ \lambda a^2}  [ \frac{4}{a^2} -   \frac{4}{3}  a  K_1 (a)    -   2 K_2 (a) + \frac{1}{3}  ]
\ee
where again $\rho$ is the instanton size and $\lambda$ the mean quark zero-mode virtuality.

After inserting (\ref{ppvertexresult}) in (\ref{ppratio1}), and using 
momentum conservation $k^\prime = p^\prime + p - k $ and $q = p - k$, 
we have
\bea\la{ppratio2}
d^{(1)}\sigma &\sim&  2^6 \frac{g_s^4}{|p-k|^4}   [ p^\prime \cdot (k + p)    \epsilon_{a \nu b c} s^a k^b p^c (p^\prime)^\nu ]  \Psi(\rho |p-k|) \nonumber\\
d^{(0)} \sigma  &\sim&  2^6 \frac{g_s^4}{|p-k|^4}  \{ 2  (k \cdot p^\prime) (p \cdot p^\prime) + (k \cdot p) [p^\prime \cdot (p - k)]                   \}  \nonumber
\eea

Using the kinematical substitution, 
\be\la{substitute}
p \longrightarrow x_1 P_1,  \ \ \  p^\prime \longrightarrow x_2 P_2,  \ \ \  k^\prime \longrightarrow \frac{K}{z}
\ee

we obtain
\bea\la{ppratio3}
d^{(1)}\sigma &\sim& 2^4 z  g_s^4 \epsilon_{a \nu b c} S^a K^b P_1^c P_2^\nu  ~ G(x_1, x_2, x_F) \nonumber\\
d^{(0)}\sigma  &\sim&    2^4 z g_s^4~ H(x_1, x_2, x_F)
\eea
with
\bea
 && G (x_1, x_2, x_F) \\
 &\equiv&   \frac{x_2}{x_1}   \frac{1 }{ ( K \cdot P_1)^2} \Psi(\rho \sqrt{   \frac{2 x_1}{z} K \cdot P_1 }~)  [ x_2 P_2  \cdot (\frac{K}{z} + x_1 P_1)  ]\nonumber
\eea

\bea
 && H (x_1, x_2, x_F)  \\
&\equiv&  \frac{x_2}{x_1} \frac{2 x_2  (K \cdot P_2)  (P_1 \cdot P_2)  +  (K \cdot P_1)  [ P_2 \cdot ( x_1 P_1 - \frac{K}{z} ) ]   }{( K \cdot P_1)^2} \nonumber
\eea

The kinematics of the collision process $p_\uparrow p$ is shown in Fig.~\ref{ppcollision3d}.

\begin{figure}
\includegraphics[height=48mm]{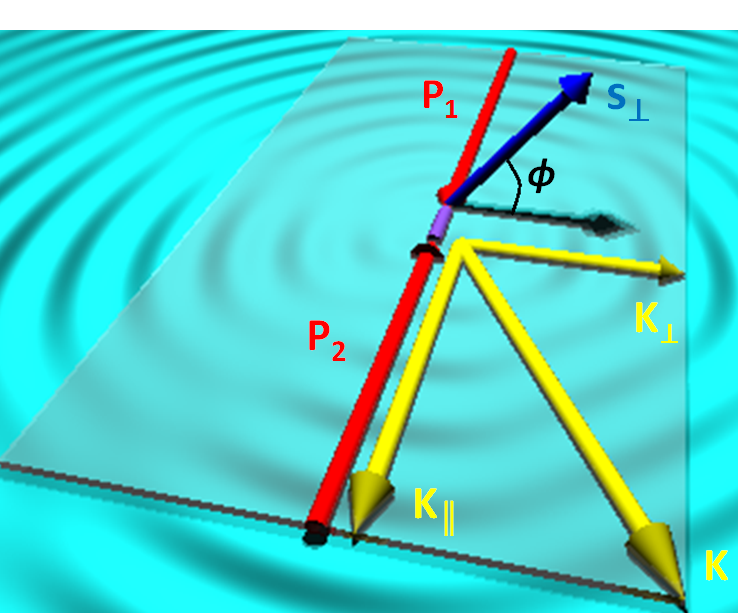}
\caption{\la{ppcollision3d} Two protons moving head-on in the CM frame. The two protons and the outgoing pion are in the same plane. The angle between the transversely polarized spin $S_\perp$ and this plane is $\phi$.  The convention for azimuthal angle is consistent with experiments.}
\end{figure}

The contribution $ \epsilon_{a \nu b c} S^a K^b P_1^c P_2^\nu $ in (\ref{ppratio3}) yields the ${\rm sin} \phi$ behavior of the asymmetry in $p_\uparrow p$ (Sivers effect)

\be
 \epsilon_{a \nu b c} S^a K^b P_1^c P_2^\nu = \frac{s}{2} K_\perp S_\perp \sin \phi
\ee
where $s = (P_1 + P_2)^2 $,  $K_\perp$ is the transverse momentum of the outgoing pion, and $S_\perp$ is transverse spin.
In the CM frame, $P_1 = \sqrt{s}/2 (1, 0 , 0 , 1)$, $P_2 = \sqrt{s}/2 (1, 0 , 0 , -1)$ and $K = (E , K_\perp, 0 , x_F \sqrt{s}/2)$, where $x_F$ is the pion longitudinal momentum fraction ($K_\parallel = x_F {\sqrt{s}}/{2}$), and $E = \sqrt{M_{\pi}^2 +K_\perp^2 + (x_F {\sqrt{s}}/{2})^2 }$ is the pion energy. $x_F>0$ stands for forward pion production, while $x_F<0$ stands for backward pion production.

Combining the above results, yields

\bea
d^{(1)}\sigma &\sim& 2^4 z  g_s^4   \frac{s}{2} K_\perp S_\perp \sin \phi ~G (x_1, x_2, x_F)     \nonumber\\
d^{(0)}\sigma  &\sim&  2^4 z g_s^4 ~  H (x_1, x_2, x_F) 
\eea

where

\begin{widetext}
\be
 G (x_1, x_2, x_F) = \frac{x_2}{x_1} \frac{\Psi(\rho \sqrt{   \frac{\sqrt{s} x_1}{z} (E - x_F \frac{\sqrt{s}}{2})} ~) }{[  \frac{\sqrt{s}}{2} (E - x_F \frac{\sqrt{s}}{2})]^2}    \{ \frac{x_2}{z} [ \frac{\sqrt{s}}{2} (E + x_F \frac{\sqrt{s}}{2}) ] + x_1 x_2 \frac{s}{2}\}
\ee

\be
 H (x_1, x_2, x_F) =  \frac{ \frac{x_2}{x_1} }{[\frac{\sqrt{s}}{2} (E - x_F \frac{\sqrt{s}}{2})]^2} \{ \frac{x_2 s^{\frac{3}{2}}}{2} (E + x_F \frac{\sqrt{s}}{2})   + \frac{\sqrt{s}}{2} (E - x_F \frac{\sqrt{s}}{2})     [ x_1\frac{s}{2} - \frac{\sqrt{s}}{2z} (E + x_F \frac{\sqrt{s}}{2}) ]    \}  
\ee
\end{widetext}

A rerun of the above analysis for the cross section arising from the process in Fig.~\ref{piplus2} yields:
\bea
d^{(1)}\sigma &\sim& - 2^4 z  g_s^4   \frac{s}{2} K_\perp S_\perp \sin \phi ~G (x_2, x_1, - x_F)     \nonumber\\
d^{(0)}\sigma  &\sim&  2^4 z g_s^4 ~  H (x_2, x_1, -x_F)
\eea

 We set $S_\perp ~ u (x, Q^2)= \Delta u_s ( x , Q^2)/2 $ and $ S_\perp ~ d (x, Q^2) = \Delta d_s ( x , Q^2)/2$, with $\Delta u_s ( x , Q^2)$ and $\Delta d_s ( x , Q^2)$ as the spin polarized distribution functions of the valence up-quarks and valence down-quarks in the proton respectively.   For simplicity, we denote $ g(x) \equiv \Delta u_s ( x , Q^2)/ u ( x , Q^2)$, $h(x) \equiv \Delta d_s ( x , Q^2)/ d ( x , Q^2)$, and $R(x)= 1/r(x) \equiv d ( x , Q^2)/u ( x , Q^2)$. Thus, the respective SSA for $\pi^+$, $\pi^-$ and $\pi^0$ follow in the form

\begin{widetext}
\bea\la{piplus}
 A_{TU}^{\sin \phi} (\pi^+) =\frac{s}{4} K_\perp \frac{g(x_1) [1 + R(x_2)] G (x_1, x_2, x_F) -[ g(x_1) + h(x_1) R(x_1) ] G (x_2, x_1, - x_F)  }{[1 + R(x_2)] H (x_1, x_2, x_F) + [1 + R(x_1)]  H (x_2, x_1, - x_F) } 
\eea

\bea\la{piminus}
 A_{TU}^{\sin \phi} (\pi^-) =\frac{s}{4} K_\perp \frac{ h(x_1) [1 + r(x_2)]  G (x_1, x_2, x_F) - [ h(x_1) + g(x_1) r(x_1) ] G (x_2, x_1, - x_F)  }{[1 + r(x_2)] H (x_1, x_2, x_F) + [1 + r(x_1)]   H (x_2, x_1, - x_F) }  
\eea

\be\la{pizero}
 A_{TU}^{\sin \phi} (\pi^0) = \frac{s}{4} K_\perp   \frac{ g(x_1)  +  R(x_1)  h(x_1) }{1 + R(x_1) } \frac{ G (x_1, x_2, x_F) - G (x_2, x_1, - x_F) }{ H (x_1, x_2, x_F) +  H (x_2, x_1, - x_F)}
\ee
\end{widetext}

The asymmetries in~(\ref{piplus}), (\ref{piminus}), and (\ref{pizero})  are propotional to $\sin \phi$. We note that for a wide
range of $x_{1,2}$  i.e. between $(0.01 - 0.99)$, $ A_{TU}^{\sin \phi} (\pi^+)$ and $ A_{TU}^{\sin \phi} (\pi^0)$ are positive while $ A_{TU}^{\sin \phi} (\pi^-)$ is negative. As the transverse momentum $K_\perp$ increases, the  asymmetries in~(\ref{piplus}), (\ref{piminus}), and (\ref{pizero}) vanish due to the modified Bessel function in the formula. To compare with the PHENIX experiment~\cite{mickey2007}, we assume the mean factorization $<x_1>=<x_2>=0.5$, $<z>=0.5$ and $<x_F>=0.3$. We have 
reset the mean instanton quark zero mode virtuality $\bar{\lambda}\approx 1/(0.2\,{\rm GeV})^3$ and instanton size $\rho \approx 0.43 $ fm.      The averaged transverse momentum of the outgoing $\pi^0$ is $<K_\perp>$ = 1 GeV. According to \cite{hirai2006} and \cite{gluck1998} ,  $g(x) = 0.959 - 0.588 (1 - x^{1.048})$, $h(x) = -0.773 + 0.478 (1 - x^{1.243})$ and  $R(x)= 1/r(x) = 0.624 (1 - x)$. In Fig.~\ref{62.4sin} we show our results  (solid line) versus the PHENIX data \cite{mickey2007}. 

\begin{figure}
\includegraphics[height=55mm]{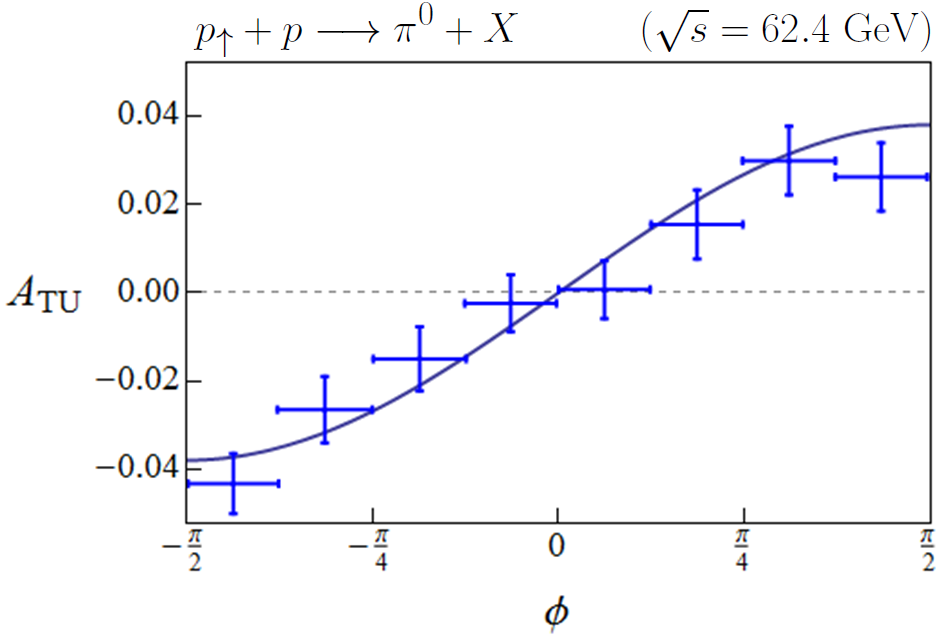}
\caption{\la{62.4sin}  $\phi$ dependent SSA in $p_\uparrow p$ collision at $\sqrt{s}$ = 62.4 GeV~\cite{mickey2007}.
Our result is the solid line.}
\end{figure}

A more thorough comparison to the wide range of kinematics swept by PHENIX and STAR $\sqrt{s}\approx (19.4-200)\, {\rm GeV}$  is displayed in Fig.~\ref{pi19.4}, Fig.~\ref{pi62.4}, and Fig.~\ref{pi200}. For simplicity, we set the parameters $<x_1>=<x_2>=0.5$ and $<z> = 0.5$ at their mean value. The instanton size is:  $\rho \approx 0.72$ fm at $\sqrt{s}$ = 19.4 GeV,  $\rho \approx 0.43$ fm at $\sqrt{s}$ = 62.4 GeV,  and  $\rho \approx 0.65$ fm at $\sqrt{s}$ = 200 GeV. Instanson size fluctuations are present in the QCD vacuum. The averaged outgoing momentum of  mesons is:  $<K_\perp> = 1$ GeV  at $\sqrt{s}$ = 19.4 GeV~\cite{fnal1991};  $<K_\perp> = 0.9$ GeV for $\eta < 3.5$ and   $<K_\perp> = 0.4$ GeV for $\eta > 3.5$  at $\sqrt{s}$ = 62.4 GeV~\cite{aidala2011}; and   $<K_\perp> = 2$ GeV for $\eta < 3.5$ and   $<K_\perp> = 1.5 $ GeV for $\eta > 3.5$ at $\sqrt{s}$ = 200 GeV~\cite{star2008}.

\begin{figure}
\includegraphics[height=58mm]{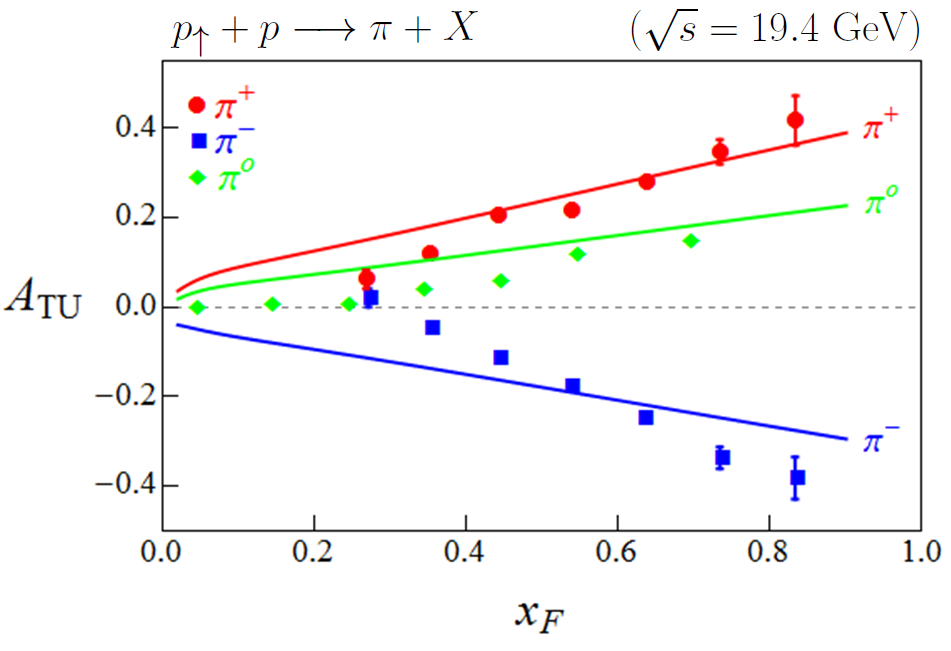}
\caption{\la{pi19.4}  $x_F$ dependent transversly polarized spin asymmetry in $p_\uparrow  p$  collision at  $\sqrt{s}$ =19.4 GeV~\cite{fnal1991}.   The averaged outgoing momentum of the pions is $<K_\perp> = 1$ GeV. The red, blue, and green solid lines stand for the SSA  in $\pi^+$ (\ref{piplus}), $\pi^-$ (\ref{piminus}) and $\pi^0$ (\ref{pizero}) productions respectively.}
\end{figure}

\begin{figure}
\includegraphics[height=95mm]{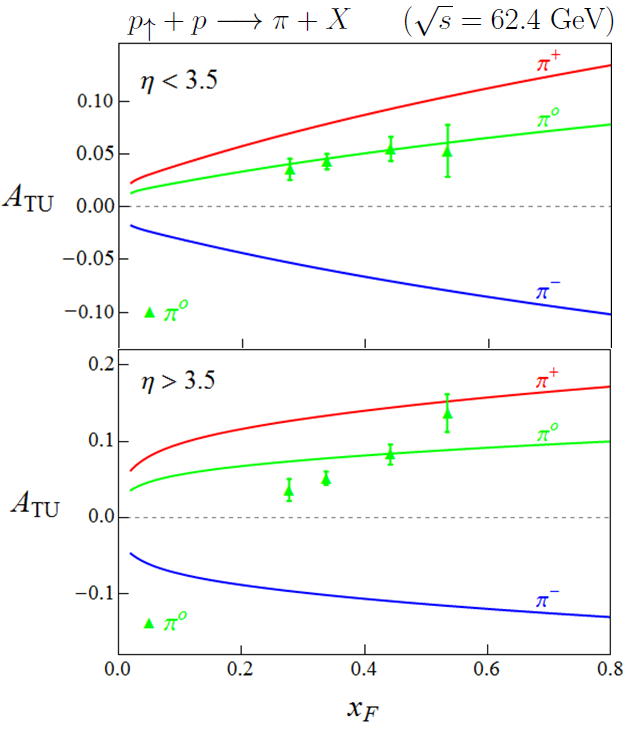}
\caption{\la{pi62.4}  $x_F$ dependent transversly polarized spin asymmetry in $p_\uparrow  p$ collision at  $\sqrt{s}$ = 62.4 GeV~\cite{aidala2011}.  The averaged outgoing momentum of $\pi^0$ is $<K_\perp> = 0.9$ GeV for $\eta < 3.5 $ and   $<K_\perp> = 0.4$ GeV for $\eta > 3.5 $. The red, blue, and green solid lines stand for the SSA  in $\pi^+$ (\ref{piplus}), $\pi^-$ (\ref{piminus}) and $\pi^0$ (\ref{pizero}) productions respectively.}
\end{figure}

\begin{figure}
\includegraphics[height=95mm]{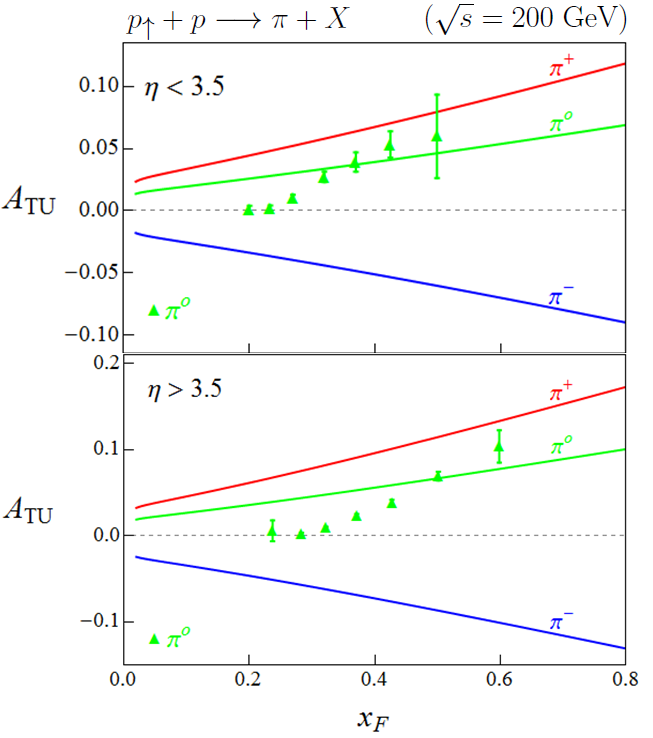}
\caption{\la{pi200}  $x_F$ dependent transversly polarized spin asymmetry in $p_\uparrow  p$ collision at  $\sqrt{s}$ =200 GeV~\cite{star2008}. The averaged outgoing momentum of $\pi^0$ is $<K_\perp> = 2 $ GeV for $\eta < 3.5 $ and   $<K_\perp> = 1.5 $ GeV for $\eta > 3.5 $. The red, blue, and green solid lines stand for the SSA in $\pi^+$ (\ref{piplus}), $\pi^-$ (\ref{piminus}) and $\pi^0$ (\ref{pizero}) productions respectively.}
\end{figure}

Recently, the SSA in backward $\pi^0$ production has been measured by the STAR collaboration~\cite{star2008}.  To compare 
our results with the data for both forward $\pi^0$ productions and  backward $\pi^0$ productions, we need to improve on the
mean factorization approximation we used with $<x_1> = <x_2> = 0.5$. This can be done by noting that
in forward $\pi^0$ production (positive $<x_F>$)  a smaller $<x_F>$ (compared to the mean factorization value of $0.5$),  
implies that $<x_2> \uparrow$ and $<x_1> \downarrow$. In contrast, a larger $<x_F>$ (compared to the mean factorization value of $0.5$), would imply $<x_2> \downarrow$ and $<x_1> \uparrow$. The same analysis reveals that $<x_1>$ and $<x_2>$ behaves oppositely for negative $<x_F>$. Therefore, we choose to parametrize the values of $<x_1>$ and $<x_2>$ as
follows: for positive $<x_F>$, $<x_1>=<x_F>+~0.2$ and  $<x_2>=0.8~-<x_F>$; for negative $<x_F>$:  $<x_2>=<x_F>+~0.2$ and  $<x_1>=0.8~-<x_F>$. The comparison is displayed in Fig.~\ref{pi200b}. Remarkably, all backward pion productions whether charged or uncharged coincide, despite the many differences in the kinematical details. This feature follows from large $\sqrt{s}$ as we now
detail.

\begin{figure}
\includegraphics[height=100mm]{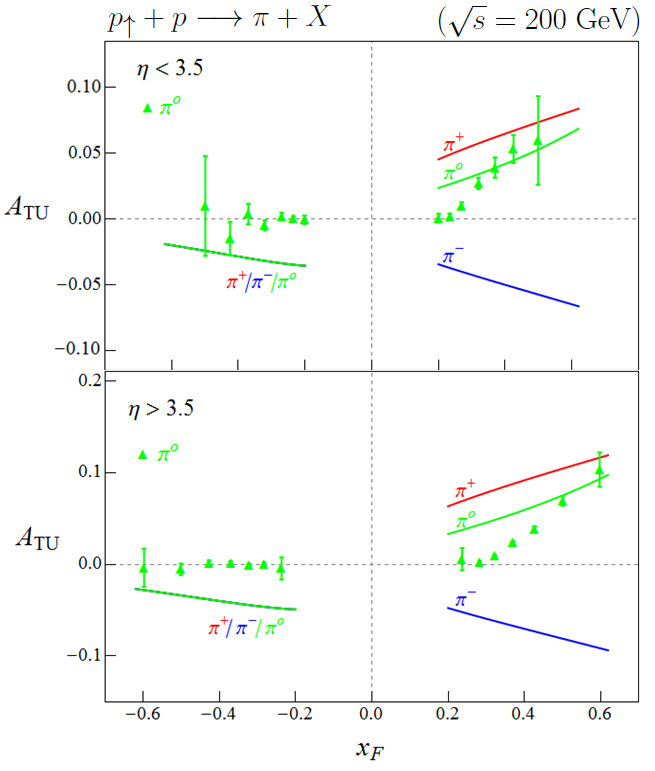}
\caption{\la{pi200b}  $x_F$ dependent transversly polarized spin asymmetry in $p_\uparrow  p$ collision at  $\sqrt{s}$ =200 GeV~\cite{star2008}. The averaged outgoing momentum of $\pi^0$ is $<K_\perp> = 2 $ GeV for $\eta < 3.5 $ and   $<K_\perp> = 1.5 $ GeV for $\eta > 3.5 $. Positive $x_f$ is forward $\pi^0$ productions. Negative $x_f$ is backward $\pi^0$ productions. The red, blue, and green solid lines stand for the SSA in $\pi^+$ (\ref{piplus}), $\pi^-$ (\ref{piminus}) and $\pi^0$ (\ref{pizero}) productions respectively. For backward pion productions ($x_F < 0$), the SSA in $\pi^+$ (\ref{piplus}), $\pi^-$ (\ref{piminus}) and $\pi^0$ (\ref{pizero}) productions coincide. }
\end{figure}

In the large $\sqrt{s}$ limit, the SSA for the three forward pion productions ($x_F>0$) simplify as only the t-channel process dominates. Specifically
\be\la{forwardplus}
 A_{TU}^{\sin \phi} (\pi^+) \approx g(x_1) \frac{ K_\perp}{4 z}   \Psi(\rho \frac{ K_\perp}{z} \sqrt{   \frac{ x_1 z }{ x_F}  }   ~) (1 +  \frac{ x_1 z }{ x_F} )
\ee
\be\la{forwardminus}
 A_{TU}^{\sin \phi} (\pi^-) \approx h(x_1) \frac{ K_\perp}{4 z} \Psi(\rho \frac{ K_\perp}{z} \sqrt{   \frac{ x_1 z }{ x_F}  }   ~) (1 +  \frac{ x_1 z }{ x_F} ) 
\ee
\bea\la{forwardzero}
 A_{TU}^{\sin \phi} (\pi^0) &\approx& \frac{g(x_1) + h (x_1) R(x_1)  }{1 + R(x_1)}
 \\
&&\times \frac{ K_\perp}{ 4 z} \Psi(\rho \frac{ K_\perp}{z} \sqrt{   \frac{ x_1 z }{ x_F}  }   ~) (1 +  \frac{ x_1 z }{ x_F} ) \nonumber
\eea

For the backward pion productions ($x_F < 0$), the u-channel process dominates leading to
\bea\la{backssa}
&& A_{TU}^{\sin \phi} (\pi^+) =  A_{TU}^{\sin \phi} (\pi^-) = A_{TU}^{\sin \phi} (\pi^0) \\
 &=& - \frac{g(x_1) + h (x_1)  R(x_1)}{1 + R(x_1)} \frac{ K_\perp}{4 z} \Psi(\rho \frac{ K_\perp}{z} \sqrt{   \frac{ x_2 z }{ |x_F|}  }   ~) (1 +  \frac{ x_2 z }{ |x_F|} ) \nonumber
\eea

For completeness, we recall that
\bea
&&g(x_1)=\frac{\Delta\,u_s(x_1, Q^2)}{u(x_1 , Q^2)}\\
&&h(x_1)=\frac{\Delta\,d_s(x_1, Q^2)}{d(x_1, Q^2)}\nonumber\\
&&\frac{g(x_1)+h(x_1)R(x_1)}{1+R(x_1)}=
\frac{\Delta u_s(x_1 , Q^2 )+\Delta d_s(x_1 , Q^2 )}{u(x_1 , Q^2 )+d(x_1 , Q^2)}\nonumber
\eea

For $\sqrt{s}>100\,{\rm GeV}$ the deviation of (\ref{forwardplus}-\ref{backssa}) from their non-asymptotic values is less than 1\%.
At large $\sqrt{s}$ the SSA in $p_\uparrow p$ following from the one-instanton insertion
is generic. All SSA for backward pion productions (\ref{backssa}) coincide as we show in Fig.~\ref{pi200b}.   In Fig.~\ref{s500} we display our predictions for the SSA for
forward pion production at $\sqrt{s} =500\, {\rm GeV} $ which is expected to be measured by the STAR collaboration. Since
our results depend on the mean value of the transverse pion momentum $<K_\perp>$ and the size of the instanton
$<\rho>$, Fig.~\ref{s500} displays a reasonable choice range. We recall that the instantons size fluctuates in the vacuum.
Overall,  our predictions are robust against this reasonable change in the parameters.

\begin{figure*}
\includegraphics[height=100mm]{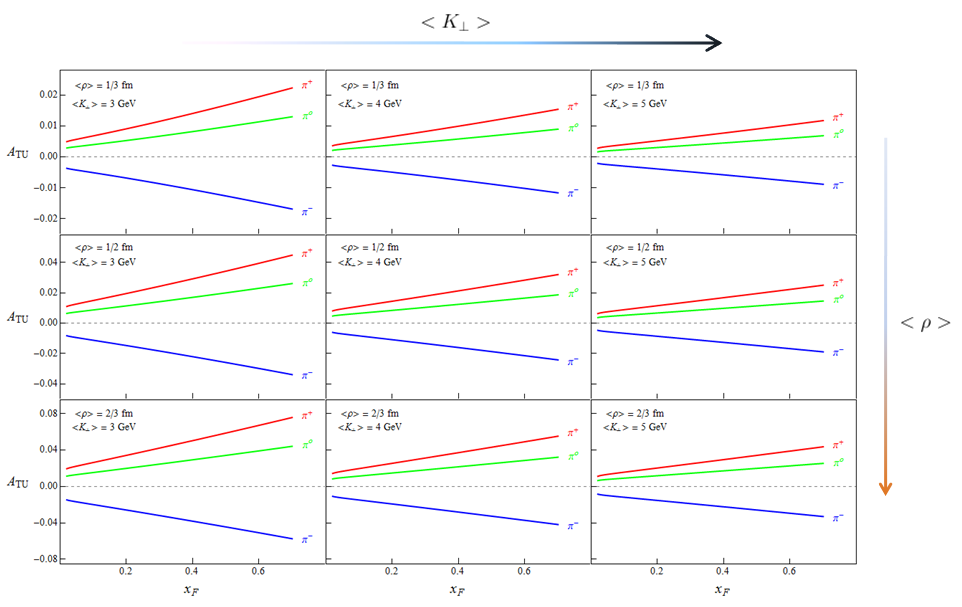}
\caption{\la{s500}  Expected $x_F$ dependence of the SSA in forward pion production in $p_\uparrow  p$ collision at  $\sqrt{s}$ =500 GeV.  }
\end{figure*}

%%%%%%%%%%%%%%%%%%%%%%%%%%%%%%
%%%%%%%%%%%%%%%%%%%%%%%%%%%%%%
\section{\label{sec:summary} Conclusions}
%%%%%%%%%%%%%%%%%%%%%%%%%%%%%%
%%%%%%%%%%%%%%%%%%%%%%%%%%%%%%

We have suggested that QCD instantons may contribute significantly to the longitudinal and transverse SSA
in SIDIS. The diluteness of the QCD instantons in the vacuum justifies the use of the single instanton approximation.
Our analysis and results are in reasonable agreement for most of the measured pion and kaon production channels. The same mechanism when applied to $p_\uparrow p$ collisions, provides a natural explanation for the Sivers contribution. Our resuts are in agreement with the reported data  in a wide range of $\sqrt{s}=19.4-200\, {\rm GeV}$. We have predicted the SSA for forward charged pion productions, and backward both charged and uncharged pion productions for a wide range of collider energies. All backward productions coincide irrespective of the differences in their kinematics at large $\sqrt{s}$. 

%%%%%%%%%%%%%%%%%%%%%%%%%%%%%%
%%%%%%%%%%%%%%%%%%%%%%%%%%%%%%
\section{\label{sec:acknowledgements}acknowledgements}
%%%%%%%%%%%%%%%%%%%%%%%%%%%%%%
%%%%%%%%%%%%%%%%%%%%%%%%%%%%%%

This work was supported in parts by the US-DOE grant DE-FG-88ER40388.

%%%%%%%%%%%%%%%%%%%%%%%%%%%%%%
%%%%%%%%%%%%%%%%%%%%%%%%%%%%%%
\section{Appendix: Photon Exchange}
\la{vertex}
%%%%%%%%%%%%%%%%%%%%%%%%%%%%%%
%%%%%%%%%%%%%%%%%%%%%%%%%%%%%%

%%%%%%%%%%%%%%%%%%%%%%%%%%%%%%
%\subsection{Photon Exchange}
%\la{dipvertex}
%%%%%%%%%%%%%%%%%%%%%%%%%%%%%%

In this Appendix, we provide the detail derivation of (\ref{dipm1}), corresponding to the nonperturbative insertion $M_\mu^{(1)}$ for photon exchange in the single instanton background. The calculation here is similar to~\cite{moch1997, ostrovsky2005}. According to~\cite{brown1978,moch1997, faccioli2001,ostrovsky2005}, the zero mode quark propagator in the single instanton background after Fourier transformation with respect to the incoming momentum $p$ is:

\be\la{simplify1}
{S_0 (x,p)_{\dot{\beta}}^{\ \ j }}_{i \delta}= \frac{2 \rho^2}{ \lambda} \frac{x^l (\overline{\sigma}_l)_{\dot{\beta} \gamma} \varepsilon^{\gamma j} \varepsilon_{i \delta}   }{(x^2 + \rho^2)^\frac{3}{2} |x|}
\ee
Note the chirality of the zero mode flips as $|L><R|$ as depicted in Fig.~\ref{flip}. The incoming quark is left-handed and has momentum $p$ (on-shell). $\rho$ is the size of instanton and $\lambda$ is the mean virtuality. $\beta$ and $\delta$ are spatial indices, while $j$ and $i$ are color indices. In Euclidean space, $\sigma_\mu = (\vec{\sigma}, i I)$, $\bar{\sigma}_\mu = (\vec{\sigma}, i I)$ and $\epsilon^{01} =-\epsilon^{10}= -\epsilon_{01} = \epsilon_{10}$~\cite{vandoren2008}.

The right-handed non-zero mode quark propagator in the single instanton after Fourier transformation with respect to the outgoing momentum $k$ is~\cite{brown1978,moch1997,ostrovsky2005}:

\bea\la{simplify2}
{S_{nz} (k, x)^{\beta i}}_{j \alpha} = &-&  [ \delta^i_j + \frac{\rho^2}{x^2} \frac{(\sigma_\rho \overline{\sigma}_r)^i_{\ \ j} k^\rho x^r}{2 k \cdot x} (1 - e^{- i k \cdot x}) ] \nonumber\\
 &\times&  \frac{|x|}{\sqrt{x^2 + \rho^2}} e^{i k \cdot x} \delta_\alpha^\beta 
\eea
As the instanton size approaches zero ($\rho \longrightarrow 0$) or the instanton is far away ($x_0 \longrightarrow \infty$), the non-zero mode propagator becomes the free propagator. This point will be revisited later.

\begin{figure}
\includegraphics[height=40mm]{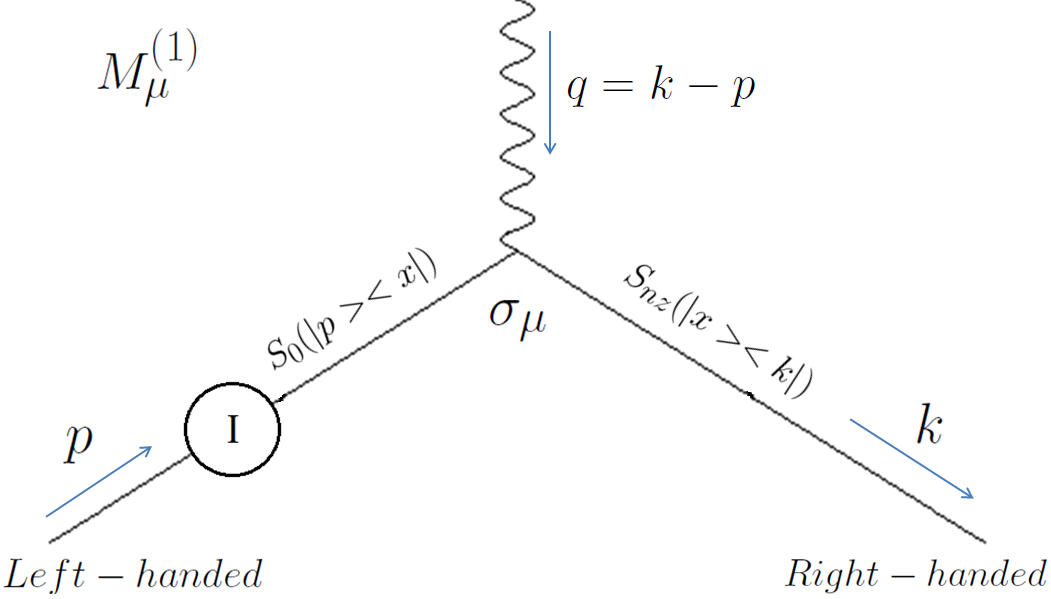}
\caption{\la{flip} The incoming left-handed quark with momentum $p$ meets one instanton and flips its chirality. The outgoing right-handed quark carries momentum $k$. The momentum of the  photon is $q = p - k$. $S_0$ and $S_{nz}$ stand for the zero-mode quark propagator and the non-zero mode quark propagator in the single instanton background respectively.  }
\end{figure}

Consider the process depicted in Fig.~\ref{flip}: the incoming left-handed quark meets one instanton and flips its chirality (zero-mode), 
then  exchanges one photon, and finally becomes an outgoing right-handed  quark. As a result, the nonperturbative insertion $M_\mu^{(1)}$ reads

\be\la{beforesimplify}
{(M_\mu^{(1)})^{\beta i}}_{i^\prime \delta} = \int d^4 x \ \  e^{- i q \cdot x} {S_{nz} (k, x)^{\beta i}}_{j \alpha} \sigma_\mu^{\alpha \dot{\beta}}{S_0 (x,p)_{\dot{\beta}}^{\ \ j }}_{i^\prime \delta}
\ee

All the other parts of the diagram are trivial in color, therefore we take the trace of color indices $i$ and $i^\prime$. 
To further simplify the result, we need the following formula~\cite{vandoren2008}

\be
  \delta_\alpha^\beta  \delta^i_j  (\sigma_\mu)^{\alpha \dot{\beta}}  (\overline{\sigma}_l)_{\dot{\beta} \gamma} \varepsilon^{\gamma j} \varepsilon_{i \delta}  = {(\sigma_\mu \overline{\sigma}_l )^\beta}_\delta
\ee

\be
  \delta_\alpha^\beta     (\sigma_\rho \overline{\sigma}_r)^i_{\ \ j}    (\sigma_\mu)^{\alpha \dot{\beta}}  (\overline{\sigma}_l)_{\dot{\beta} \gamma} \varepsilon^{\gamma j} \varepsilon_{i \delta} = {(\sigma_\mu \overline{\sigma}_l \sigma_r \overline{\sigma}_\rho)^\beta}_\delta
\ee

Combining all the equations above, we obtain
\bea\la{combine1}
M_\mu^{(1)} =&-& \int d^4 x \ \  [ \frac{2 \rho^2}{ \lambda}   \sigma_\mu \overline{\sigma}_l   e^{ i p \cdot x} \frac{x^l}{(x^2+ \rho^2)^2} \\
&+&  \sigma_\mu \overline{\sigma}_\rho k^\rho \frac{ \rho^4}{\lambda}( e^{i p \cdot x} - e^{- i q \cdot x} )  \frac{1}{(x^2+ \rho^2)^2 ( k \cdot x)}  ]  \nonumber
\eea

The $d^4 x$ integration in (\ref{combine1}) can be done with the help of the following formula ($p^2 \longrightarrow 0$)

\be\la{dipvanish}
\int d^4 x \ \  e^{ i p \cdot x} \frac{x^l}{(x^2+ \rho^2)^2} =  i 2 \pi^2 \frac{ p^l}{p^2}
\ee
\bea\la{formula1}
\int d^4 x    \frac{  e^{ i p \cdot x}}{(x^2+ \rho^2)^2 ( k \cdot x)} &=&  -i \frac{2 \pi^2}{p \cdot k} \frac{\rho |p|}{\rho^2} K_1  (\rho |p|)  \nonumber\\
&=& i \frac{4 \pi^2}{q^2} \frac{\rho |p|}{\rho^2} K_1  (\rho |p|) 
\eea
where we used $- 2 p \cdot k = (k - p)^2 -k^2 - p^2 \approx q^2$. The demoninator in (\ref{dipvanish}) appears to diverge 
as $p^2 \longrightarrow 0$. We show below that all terms proportional to $\overline{\sigma}_l p^l$ vanish.  

Thus

\be\la{dipfinal1}
M_\mu^{(1)} =-  i \frac{4 \pi^2 \rho^2}{\lambda} \sigma_\mu  \overline{\sigma}_l [ \frac{k^l}{q^2} (f(\rho |p|) - f(\rho |q|)) +  \frac{p^l}{p^2} ]
\ee
where $f(a) = a K_1 (a)$
As the incoming quark with momentum $p$ is on-shell and the mass of the quark is small ($p^2 \longrightarrow 0$), we have

\be
f(\rho |p|) = \rho |p| K_1 (\rho |p|) \longrightarrow \rho |p| \frac{1}{\rho |p|} = 1
\ee
Since $q^2 <0$ in SIDIS, we define $Q^2 = - q^2 > 0$. (\ref{dipfinal1}) simplifies to

\be\la{dipfinal2}
M_\mu^{(1)} = i \frac{4 \pi^2 \rho^2}{\lambda} \sigma_\mu  \overline{\sigma}_l  [ \frac{k^l}{Q^2} (1 - f(\rho Q)) -  \frac{p^l}{p^2}]  
\ee

If we consider that the incomging right-handed quark flips its spin and becomes left-handed one, we only need to conjugate 
(\ref{dipfinal2}) and substitute $k \leftrightarrow - p$

\bea\la{dipfinal3}
M_\mu^{(1)} = &&  i \frac{4 \pi^2 \rho^2}{\lambda}( \sigma_\mu  \overline{\sigma}_l k^l + \sigma_l  \overline{\sigma}_\mu p^l ) \frac{1}{Q^2} (1 - f(\rho Q)) \nonumber\\
&& -   i \frac{4 \pi^2 \rho^2}{\lambda}( \sigma_\mu  \overline{\sigma}_l \frac{p^l}{p^2} + \sigma_l  \overline{\sigma}_\mu \frac{k^l}{k^2} )
\eea
If the anti-instanton is also included ($\sigma \leftrightarrow \overline{\sigma}$), we have the compact form

\bea\la{dipfinal4}
M_\mu^{(1)} = &&  i \frac{4 \pi^2 \rho^2}{\lambda Q^2}[ \gamma_\mu \slashed{k} + \slashed{p} \gamma_\mu] (1 - f(\rho Q))  \nonumber\\
&& -   i \frac{4 \pi^2 \rho^2}{\lambda} [\gamma_\mu \frac{\slashed{p}}{p^2} + \gamma_\mu \frac{\slashed{k}}{k^2}]
\eea
The last two terms in (\ref{dipfinal4}) vanish. Indeed, set  $\gamma_\mu {\slashed{p}}/{p^2}$ and insert it in the hadronic tensor
$W_{\mu\nu}$ (\ref{HADRONIC}). Thus

\bea
\lim_{p^2 \rightarrow 0} \tr [\gamma_\mu \frac{\slashed{p}}{p^2} \slashed{p} \gamma_5 \slashed{s} \gamma_\nu \slashed{k}] = \tr [\gamma_\mu \gamma_5 \slashed{s} \gamma_\nu \slashed{k}]
\eea
where the indices $\mu$ and $\nu$ are anti-symmetric. The same indices are symmetric in the leptonic tensor (\ref{LEPTONIC}). 
Thus the two contract to zero.

Finally, we explicitly show that all terms proportional to $\overline{\sigma}_l p^l$ in (\ref{dipfinal1}) vanish. For that, rewrite
(\ref{simplify2}) as

\bea\la{combine2}
S_{nz} (k, x) = &-&   \frac{\rho^2}{x^2} \frac{(\sigma_\rho \overline{\sigma}_r)^i_{\ \ j} k^\rho x^r}{2 k \cdot x} \frac{ (1 - e^{- i k \cdot x}) |x|}{\sqrt{x^2 + \rho^2}} e^{i k \cdot x} \delta_\alpha^\beta  \nonumber\\
 &-& \frac{|x|}{\sqrt{x^2 + \rho^2}} e^{i k \cdot x} \delta_\alpha^\beta  \delta^i_j 
\eea
As the instanton size $\rho\rightarrow 0$, the first contribution in (\ref{combine2}) vanishes and the second contribution reduces to the free propagator. As a result, the chirality flip vertex $M_\mu^{(1)}$ vanishes. The second contribution in (\ref{combine2}) is at the
origin of the term proportional $\overline{\sigma}_l p^l$. This analysis is general and will be exported to the gluon vertex below.

To summarize: the instanton vertex to be used following a photon insertion is

\be\la{dipfinal5}
M_\mu^{(1)} =  i \frac{4 \pi^2 \rho^2}{\lambda Q^2}[ \gamma_\mu \slashed{k} + \slashed{p} \gamma_\mu] (1 - f(\rho Q)) 
\ee
which is (\ref{dipm1}).

%%%%%%%%%%%%%%%%%%%%%%%%%%%%%%
\section{Appendix: Gluon Exchange}
\la{ppvertex}
%%%%%%%%%%%%%%%%%%%%%%%%%%%%%%

In this Appendix we derive (\ref{ppvertexresult}), corresponding to the nonperturbative insertion $M_\mu^{(1)}$ for gluon exchange in the single instanton background. The calculation is similar to the photon exchange (\ref{vertex}). 
Consider  the following case: the incoming left-handed quark with momentum $p$ flips its chirality (zero-mode), exchange one gluon, and finally becomes the right-handed outgoing quark with momentum $k$ in the single instanton background as depicted in
Fig.~\ref{ppflip}.

\begin{figure}
\includegraphics[height=40mm]{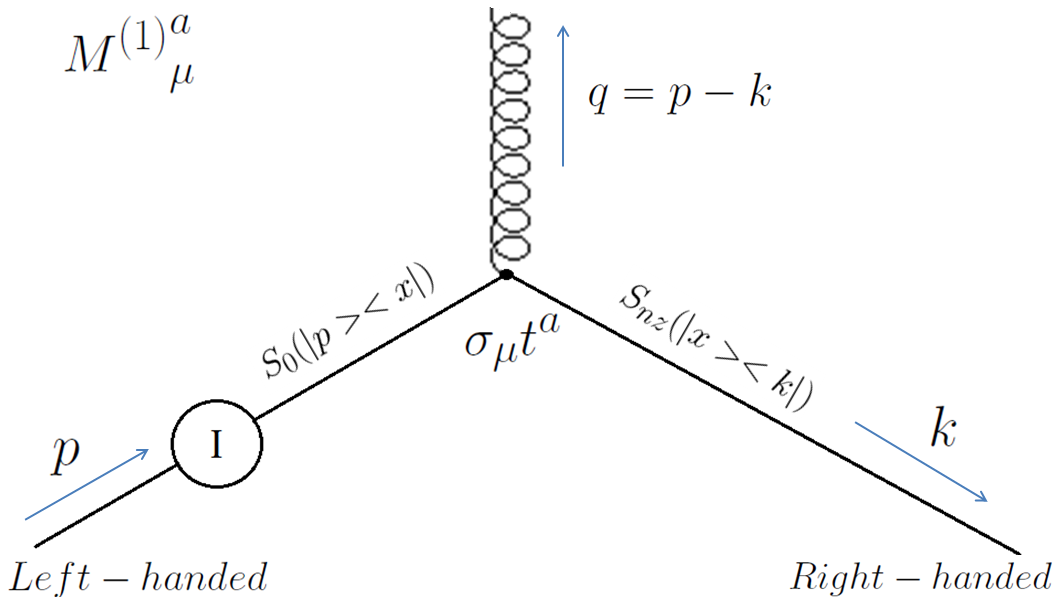}
\caption{\la{ppflip} The incoming left-handed quark with momentum $p$ meets one instanton and flips its chirality. The outgoing right-handed quark carries momentum $k$. The momentum of the gluon is $q = k - p$.  $S_0$ and $S_{nz}$ stand for zero-mode quark propagator and non-zero mode quark propagator in single instanton background respectively. }
\end{figure}

This diagram is similar to the photon exchange one, but now the gluon carries color. Since the instanton is 
SU(2) valued, we embed the SU(2) in the upper corner of SU(3) to perform the color tracing.  Thus

\bea\la{embedzero}
{S_0 (x,p)_{\dot{\beta}}^{\ \ j }}_{i \delta} = 
\begin{cases}
 \frac{2 \rho^2}{ \lambda} \frac{x^l (\overline{\sigma}_l)_{\dot{\beta} \gamma} \varepsilon^{\gamma j} \varepsilon_{i \delta}   }{(x^2 + \rho^2)^\frac{3}{2} |x|}   \ \ \ \  &(i,j = 1,2)\\
0 &  (j=3 \ \  or \ \  i=3)
\end{cases}
\eea
where $S_0 (x,p)$ is the zero mode quark propagator in the single instanton background after Fourier transforming with 
respect to the incoming particle momentum $p$.

\bea\la{embednonzero}
&& {S_{nz} (k , x)^{\beta i}}_{j \alpha}   =\\
&& 
\begin{cases}
- \frac{|x|}{\sqrt{x^2 + \rho^2}} e^{i k \cdot x} \delta_\alpha^\beta [ \delta^i_j + \frac{\rho^2}{x^2} \frac{(\sigma_\rho \overline{\sigma}_r)^i_{\ \ j} k^\rho x^r}{2 k \cdot x}  (1 - e^{- i k \cdot x}) ] \\  
\ \ \ \ \ \ \ \ \ \ \ \ \ \ \ \ \ \ \ \ \ \ \ \ \ \ \ \ \ \ \ \ \ \ \ \ \ \ \ \  (i , j =1,2)  \\
 - e^{i k \cdot x} \delta^\beta_\alpha \delta^i_j  \ \ \ \ \ \ \ \ \ \ \ \ \ \ \ \ \ \ \ \ \ \ \ \ \ \ \ \ \     (i=3)    \nonumber
\end{cases}
\eea
where $S_{nz} (k , x)$ is the right-handed non-zero mode quark propagator in the single instanton after Fourier transforming with
respect to outgoing particle with momentum $k$.

The nonperturbative insertion $ M^{(1)}$ in (\ref{ppratio1}) is given by
\be\la{ppm1}
{M^{(1)}}_\mu^a  = \int d^4 x \ \  e^{ i q \cdot x}{S_{nz} (k, x)^{\beta i}}_{j \alpha} \sigma_\mu^{\alpha \dot{\beta}} (t^a)^j_{\ \ j^\prime} {S_0 (x,p)_{\dot{\beta}}^{\ \  j^\prime }}_{i^\prime \delta}
\ee

Combining (\ref{embedzero}),  (\ref{embednonzero}) and (\ref{ppm1}),  we get (\ref{ppvertexresult}). To keep the number of indices
from mushrooming we need to recombine the expressions. The first observation is that the non-zero mode  $S_{nz} (k , x)$ will become the free propogator if the instanton size $\rho \longrightarrow 0$. Therefore, we first pick the terms with trial color indices in ${S_{nz} (k, x)^{\beta i}}_{j \alpha}$ (these contain $\delta^i_j$). The contribution to the vertex of such terms is given by

\bea\la{zero1}
&&- \frac{2 \rho^2}{\lambda} \delta^{ab} \int d^4 x \ \ e^{i (k + q) \cdot x} \frac{x^l (\sigma_\mu \overline{\sigma}_l \epsilon)^{\beta j^\prime}}{(x^2 + \rho^2)^{\frac{3}{2}}|x|} \\
&\times& \{ {[t_a^T]_{j^\prime}}^{i=3}  {[t_b^T]_{i=3}}^{i^\prime}   + \sum_{i=1,2}\frac{|x|}{\sqrt{x^2+ \rho^2}} \nonumber {[t_a^T]_{j^\prime}}^{i}  {[t_b^T]_{i}}^{i^\prime} \} \epsilon_{i^\prime \delta}
\eea
Since $k + q = p$ and $p^2 \approx 0$, we can simplify (\ref{zero1}) into
\bea
&&- \frac{2 \rho^2}{\lambda}  \int d^4 x \frac{e^{i p \cdot x}}{|x|} \frac{x^l (\sigma_\mu \overline{\sigma}_l \epsilon)^{\beta j^\prime}}{(x^2 + \rho^2)^{\frac{3}{2}}}  \sum_{i=1,2,3} {[t_a^T]_{j^\prime}}^{i}  {[t^a]^T}_i^{i^\prime}   \epsilon_{i^\prime \delta} \nonumber\\
&=&  -\frac{8 \rho^2}{3 \lambda} (\sigma_\mu \overline{\sigma}_l )  \int d^4 x \ \ e^{i p \cdot x} \frac{x^l}{x^2} \nonumber\\
&\longrightarrow& (\cdots)\overline{\sigma}_l  p^l
\eea
As we discussed earlier, the terms propotional to $\overline{\sigma}_l  p^l$ finally vanish (see (\ref{vertex})). 

Collecting all the terms left, we have

\bea\la{ppleft}
&& - \frac{ \rho^4}{\lambda} \delta^{ab} \int d^4 x \ \ \ (e^{i p \cdot x} - e^{ i q \cdot x}) \frac{1}{(x^2 + \rho^2)^2} \frac{1}{x^2} \frac{k^\rho x^r x^l}{ k \cdot x} \nonumber\\
 &\times& [  \sum_{i=1,2} \sum_{j=1,2}  (\sigma_\mu \overline{\sigma}_l \epsilon)^{\beta j^\prime} [t_a^T]_{j^\prime}^{\ \ j} { [\sigma_\rho \overline{\sigma}_r]^T}_j^{\ \ i} [t_b^T]_i^{ \ \ i^\prime} \epsilon_{i^\prime \delta}] \nonumber\\
\eea
Note that $\{ i , j\} = \{1 ,2 \}$ is due to our choice of the SU(2) instanton embedding into $SU(3)$. The final result should not dependent on this embedding. Indeed, for an arbitrary SU(2) group element $g$ we expect

\bea
  \sum_{i,j=\{ 1,2\} } \delta^{a b}  [t_a^T]_{j^\prime}^{\ \ j} { g}_j^{\ \ i} [t_b^T]_i^{ \ \ i^\prime} &=&   \sum_{i,j=\{2,3\} }  \delta^{a b}  [t_a^T]_{j^\prime}^{\ \ j} { g}_j^{\ \ i} [t_b^T]_i^{ \ \ i^\prime} \nonumber\\
 &=&   \sum_{i,j=\{ 3,1\} } \delta^{a b}   [t_a^T]_{j^\prime}^{\ \ j} { g}_j^{\ \ i} [t_b^T]_i^{ \ \ i^\prime} \nonumber
\eea
The coefficients are given by

\be\la{co1}
  \sum_{i,j=\{ 1,2\} } \delta^{a b}  [t_a^T]_{j^\prime}^{\ \ j}  { I }_j^{\ \ i} [t_b^T]_i^{ \ \ i^\prime} = \frac{5}{6} {I_{j^\prime}}^{i^\prime}
\ee

\be\la{co2}
  \sum_{i,j=\{ 1,2\} } \delta^{a b}  [t_a^T]_{j^\prime}^{\ \ j}  {( \sigma_m )}_j^{\ \ i} [t_b^T]_i^{ \ \ i^\prime} = -  \frac{1}{6} {(\sigma_m)_{j^\prime}}^{i^\prime}
\ee
These two coefficients are different as the instaonton color indices are diagonal and couple with spin in a non-trivial way.

Inserting (\ref{co1}) and (\ref{co2}) in ({\ref{ppleft}}), we obtain new terms. The $d^4 x$ integration can be done with the help of (\ref{dipvanish}), (\ref{formula1}) and

\bea
&&    \int d^4 x \ \ \ e^{  i q \cdot x} \frac{1}{(x^2 + \rho^2)^2} \frac{x^l}{x^2} \\
&=&  i  2 \pi^2  q^l \ \  \frac{4 - (\rho |q| )^3 K_1 ( \rho |q| ) - 2 (\rho |q|)^2 K_2 (\rho |q| )}{ (\rho |q|)^4}  \nonumber
\eea

Combining the previous results, substituting $q = p - k$ and $|q| = \sqrt{2 p \cdot k}$, and dropping all the terms propotional to $\overline{\sigma}_l p^l$,  yield

\be
(M^{(1)})_\mu^a t^b \delta_{a b} = -i  \sigma_\mu \overline{\sigma}_l k^l    \Psi(\rho |q|) 
\ee

with
\be\la{ppfinal1}
\Psi(a)  \equiv \frac{ 2 \pi^2 \rho^4}{ \lambda a^2}  [ \frac{4}{a^2} -   \frac{4}{3}  a  K_1 (a)    -   2 K_2 (a) + \frac{1}{3}  ]
\ee
Note that if instead a right-handed incoming quark flips its chirality and becomes a left-handed outgoing quark, we only need to conjugate (\ref{ppfinal1}) and substitute $ k \longrightarrow - p$. If an instanton is switched to an 
anti-instanton, then $\sigma \leftrightarrow \overline{\sigma}$ in the final result. With this in mind, we have

\be
M_\nu^a t^b \delta_{a b} = -i  (\gamma_\mu \slashed{k} + \slashed{p} \gamma_\mu)   \Psi(\rho |q|) 
\ee
which is (\ref{ppvertexresult}).

\bibliography{instantonref}

\end{document}